\documentclass[11pt]{article} % font size
\usepackage[utf8]{inputenc}
\usepackage[english]{babel}
\usepackage[a4paper,margin=1in]{geometry} % page setup, margins
\usepackage{authblk} % author affiliations
\usepackage[colorlinks=true, linkcolor=black, citecolor=black, urlcolor=blue, filecolor=blue,breaklinks=true]{hyperref} % Hyperlinks, needed for biblatex, other option: allcolors=blue
\usepackage{setspace} % for linespacing
\usepackage{lineno} % line numbering (change with \linespacing below)
\usepackage{csquotes} % Recommended for biblatex, must load after lineno or will get a warning
\usepackage{ragged2e} % for left aligning
\usepackage[backend=biber, natbib=true, sorting=none]{biblatex}
\usepackage{siunitx}
\usepackage{mhchem}

\usepackage[capitalise, noabbrev, nameinlink]{cleveref}
\crefdefaultlabelformat{#2\textbf{#1}#3} % To bold figure refs
\Crefname{figure}{\textbf{Figure}}{\textbf{Figures}} % To bold figure refs
\Crefname{section}{\textbf{Section}}{\textbf{Sections}} % To bold section refs
\Crefname{table}{\textbf{Table}}{\textbf{Tables}} % To bold table refs
\makeatletter
\newcommand*{\rom}[1]{\expandafter\@slowromancap\romannumeral #1@}
\makeatother
\usepackage{graphicx}
\graphicspath{{./main_figs/}{./supplement/suppl_figs/}} % paths for figs
\DeclareGraphicsExtensions{.pdf,.jpeg,.JPG,.png,.PNG, .eps, .tiff}
\usepackage{subcaption} % for subcaptions (panels), and add parentheses
\DeclareCaptionLabelFormat{bold}{\textbf{(#2)}} % bold subpanel letter in caption
\newcommand{\labelphantom}[1]{%  To make subpanel references easier from https://tex.stackexchange.com/a/255790/121424 
  \parbox{0pt}{\phantomsubcaption\label{#1}}%
}
\usepackage{pgffor} % for function to make figures with subpanels
\usepackage{alphalph} % for function to make figures with subpanels
\usepackage[labelfont=bf, textfont=bf, singlelinecheck=off, textfont=footnotesize]{caption} % boldness of captions
\usepackage{booktabs}
\usepackage{multirow}
\usepackage{rotating}
\newcommand{\generateFigSubpanels}[6][0]{% default rotation=0
     \expandafter\newcommand\csname#2\endcsname{ %https://tex.stackexchange.com/questions/65780/macro-defining-macro/65781#65781
      \begin{figure}[htbp]
        \foreach [count=\i] \x in #6{%
            \labelphantom{fig:#2:\AlphAlph{\i}}
        }
        \centering
        \includegraphics[width=#4\linewidth,angle=#1]{#3}
        \caption{\textbf{\normalsize #5}}\label{fig:#2}
        \footnotesize
        \justifying
        \foreach [count=\i] \x in #6{%
            \noindent\subref{fig:#2:\AlphAlph{\i}}~\x\space
        }
      \end{figure}
    }
}

\newcommand{\generateFig}[6][0]{%  default rotation=0
     \expandafter\newcommand\csname#2\endcsname{ %https://tex.stackexchange.com/questions/65780/macro-defining-macro/65781#65781
      \begin{figure}[htbp]
        \centering
        \includegraphics[width=#4\linewidth,angle=#1]{#3}
        \caption{\textbf{\normalsize #5}\label{fig:#2}
        \footnotesize
        \normalfont
        #6
        }
        
      \end{figure}
    }
}

\newcommand{\generateSidewaysFigSubpanels}[6][0]{% default rotation=0
     \expandafter\newcommand\csname#2\endcsname{ %https://tex.stackexchange.com/questions/65780/macro-defining-macro/65781#65781
      \begin{sidewaysfigure}[htbp]
        \foreach [count=\i] \x in #6{%
            \labelphantom{fig:#2:\AlphAlph{\i}}
        }
        \centering
        \includegraphics[width=#4\linewidth,angle=#1]{#3}
        \caption{\textbf{\normalsize #5}}\label{fig:#2}
        \footnotesize
        \justifying
        \foreach [count=\i] \x in #6{%
            \noindent\subref{fig:#2:\AlphAlph{\i}}~\x\space
        }
      \end{sidewaysfigure}
    }
}

\newcommand{\generateSidewaysFig}[6][0]{%  default rotation=0
     \expandafter\newcommand\csname#2\endcsname{ %https://tex.stackexchange.com/questions/65780/macro-defining-macro/65781#65781
      \begin{sidewaysfigure}[htbp]
        \centering
        \includegraphics[width=#4\linewidth,angle=#1]{#3}
        \caption{\textbf{\normalsize #5}\label{fig:#2}
        \footnotesize
        \normalfont
        #6
        }
        
      \end{sidewaysfigure}
    }
}

\newcommand{\generateTab}[6][]{%  Note that rotation does not work here
     \expandafter\newcommand\csname#2\endcsname{ \begin{table}[ht]
          \centering
                \resizebox{#4\textwidth}{!}{
                \input{#3}
            }
            \caption{\textbf{\normalsize #5}\label{tab:#2}
            \footnotesize
            \normalfont
            #6
            }
            
        \end{table}
    }
}

\newcommand{\generateSidewaysTab}[6][]{%  Note that rotation does not work here
     \expandafter\newcommand\csname#2\endcsname{ \begin{sidewaystable}[ht]
          \centering
                \resizebox{#4\textwidth}{!}{
                \input{#3}
            }
            \caption{\textbf{\normalsize #5}\label{tab:#2}
            \footnotesize
            \normalfont
            #6
            }
            
        \end{sidewaystable}
    }
}

\usepackage[dvipsnames]{xcolor} % can remove, for coloring the tips
\usepackage{changepage} % can remove, for width on the tips
\usepackage{enumitem} % can remove, for width on the tips

\title{Dense and porous phase transition study of SEI formation using phase-field method}

\author[1]{Joonyeob Jeon}
\author[2,3]{Lukas Köbbing}
\author[1]{Tejs Vegge}
\author[1]{Jin Hyun Chang}
\author[2,3,4,*]{Birger Horstmann}
\author[1,*]{Ivano E. Castelli}
\affil[1]{Department of Energy Conversion and Storage, Technical University of Denmark, Kongens Lyngby, Denmark}
\affil[2]{Institute of Engineering Thermodynamics, German Aerospace Center (DLR), Wilhelm-Runge-Straße 10, 89081 Ulm, Germany}
\affil[3]
{Helmholtz Institute Ulm (HIU), Helmholtzstraße 11, 89081 Ulm, Germany}
\affil[4]{Faculty of Natural Sciences, Ulm University, Albert-Einstein-Allee 47, 89081 Ulm, Germany}

\affil[ ]{ } % for some blank space
\affil[*]{\textit{Correspondence to }\underline{ivca@dtu.dk} and \underline{birger.horstmann@dlr.de}} 
\date{} % If you want a date fill this in (e.g. \date{February 2022})

\newcommand{\makeAbstract}{
\begin{abstract}
The solid electrolyte interphase (SEI) is essential for the long-term stability of batteries because it influences the reactions between the electrode and electrolyte. Despite previous studies, the SEI evolution from dense to porous phases remains incompletely understood. Here, we investigate the SoC-dependent evolution of the SEI using a phase-field framework under open-circuit conditions. Spatially correlated noise is introduced to describe stochastic transport perturbations, and the dense-to-porous transition time is evaluated from the evolution of interface roughness arising from the competition between noise-induced transport and surface relaxation. The simulations reveal three distinct roughness evolution regimes. The predicted transition time generally increases from approximately one month at \SI{20}{\%} SoC to more than seven months at \SI{80}{\%} SoC, with a pronounced change between \SI{55}{\%} and \SI{60}{\%}. This change corresponds to the graphite SoC--OCV relation, which controls the interfacial $\ce{Li^0}$ radical concentration through the Nernst condition. Under dynamic SoC conditions caused by irreversible capacity loss, the transition behavior changes as the SoC evolves during storage. These results show that the SoC--OCV relation is central to determining SEI phase stability and the dense-to-porous transition time under open-circuit storage.
\end{abstract}

}

\begin{document}

    %\RaggedRight % If you want left-aligned (ie ragged right text) rather than justified
    
    \maketitle
    \makeAbstract
    \clearpage 
    % --- Task 1.1: Establish SEI Context & Complexity ---
% Goal: Define SEI and explain why it is a subject of intense interest.
\section{Introduction}
The Solid Electrolyte Interphase (SEI) is a passivating film that forms between the electrode and the electrolyte during battery cycling. SEI formation is observed not only in commercial Li-ion batteries with graphite-based anode \cite{an_state_2016, gomez_rojas_electrolyte_2025, shi_defect_2013, strmcnik_electrocatalytic_2018, qin_computational_2024} but also in many other battery chemistries, such as solid-state \cite{wu_prediction_2022, ren_visualizing_2024, spotte-smith_toward_2022}, \ce{Li} metal \cite{von_holtum_primary_2025, ren_visualizing_2024, shan_brief_2021}, and \ce{Si} anode \cite{soto_formation_2015, soto_understanding_2018,philippe_nanosilicon_2012, philippe_role_2013, xu_improved_2015}. Although its formation results from electrolyte decomposition, the formed SEI substantially suppresses further electrolyte decomposition and protects the electrode surface, thereby directly affecting battery performance, stability, and lifetime \cite{wang_review_2018, heiskanen_generation_2019, spotte-smith_toward_2022, an_state_2016}. Nevertheless, the SEI has remained a subject of sustained investigation for decades because of its complex chemical composition, spatially heterogeneous structure, and the analytical difficulty posed by the simultaneous reactions of multiple chemical and electrochemical processes \cite{peled_electrochemical_1979, kobbing_growth_2023, wang_review_2018}. In particular, under open-circuit storage or long-term aging conditions, SEI growth can continue even in the absence of an external current \cite{bieker_electrochemical_2015}, and this continued growth leads to irreversible lithium loss and capacity fade. Therefore, understanding and predicting SEI growth behavior, particularly in graphite anodes that constitute the dominant commercial Li-ion chemistry \cite{zhao_progress_2024}, is essential for evaluating the long-term performance and stability of lithium-ion batteries.

% --- Task 1.2: Introduce the Composition of SEI (Bilayer Structure) ---
% Goal: Categorize SEI into Organic and Inorganic layers.

The long-term stability of the SEI is governed by multiple factors, including chemical composition \cite{philippe_nanosilicon_2012, philippe_role_2013, xu_improved_2015}, ionic and electronic transport \cite{soto_understanding_2018, shi_defect_2013}, mechanical properties \cite{zheng_3d_2014, shang_lattice_2012, shin_component-structure-dependent_2015}, and microstructural state evolution \cite{von_kolzenberg_transition_2022, single_identifying_2018, lu_chemistry_2014}. Among these factors, the microstructural state of the SEI is particularly important because it determines the extent of electrolyte penetration and the reactive interfacial area \cite{single_revealing_2017, an_state_2016}. A dense SEI can effectively cover the electrode surface and significantly reduce the rate of further electrolyte decomposition, whereas a porous SEI provides pathways for electrolyte penetration and promotes additional side reactions \cite{heiskanen_generation_2019}. Therefore, evaluating the dense-to-porous transition is important for understanding how SEI growth affects long-term interfacial stability. 

The dense and porous structures of the SEI have generally been interpreted in relation to the layered arrangement of organic and inorganic species, with inorganic-rich domains typically located closer to the electrode and organic-rich domains more commonly present toward the electrolyte side of the SEI \cite{philippe_nanosilicon_2012, an_state_2016, wang_review_2018, xu_improved_2015}. This perspective has recently been extended to simulation studies by considering how the different material properties of dense and porous SEI regions affect SEI growth and transport.

\citet{zhang_simulating_2025} employed multiphase-field modeling to predict the formation of organic and inorganic SEI phases during the early stage of battery cycling. This approach illustrates how phase-field modeling can describe the coupled evolution of multiple SEI phases within a continuum framework. More generally, the phase-field method provides a numerical framework for describing interfacial evolution without explicitly tracking the moving boundary. This feature has made phase-field modeling a useful tool for studying electrochemical and nanostructured systems, including SEI and battery interface evolution \cite{liang_nonlinear_2014, bazant_theory_2013, chen_modulation_2015, cogswell_quantitative_2015, hong_phase-field_2018, zhang_quantitative_2023}.
Meanwhile, the dense-to-porous transition of the SEI has also been proposed to arise from transport-limited growth dynamics, even in an SEI consisting of a single chemical species. \citet{single_revealing_2017} derived an analytical growth law showing that the SEI thickness follows a $\sqrt{t}$ dependence. In this framework, SEI growth is assumed to be limited by electron diffusion rather than by interfacial reaction kinetics, and this transport-limited description was shown to capture the experimentally observed long-term growth pattern of the SEI \cite{kobbing_growth_2023, single_identifying_2018}. Extending this framework to the problem of morphological transition, \citet{von_kolzenberg_transition_2022} combined 1D electron diffusion, barrier-height variation, and stochastic noise to model the dense-to-porous transition of the SEI and to predict dual-layer SEI and SoC-dependent transition times. 

These studies provide an important foundation for describing SEI formation, long-term growth, and dense-to-porous transition. Three points remain insufficiently resolved. First, the dense-to-porous transition has mainly been described using analytical or 1D frameworks. These approaches are useful for capturing long-term growth trends and transition times, but they do not directly resolve how local reaction rates change along an evolving SEI surface in a 2D domain.
Second, the transition from dense to porous SEI occurs gradually rather than abruptly, yet the pattern of this gradual transition remains insufficiently understood. The influence of the SEI surface has been less extensively explored due to the limitations of 1D simulation.
Third, analytical models have predicted dense-to-porous SEI transitions and their dependence on the SoC. However, the evolution of the SEI phase under coupled reaction and transport remains unresolved, particularly with respect to the effects of time-dependent SoC reduction. Consequently, it is still unknown how the SEI phase evolves toward a porous state under open-circuit aging conditions, where the SoC gradually decreases over time due to irreversible capacity loss.

Here, we develop a phase-field framework to analyze the dense-to-porous transition of the SEI under transport-driven stochastic instability, applied to a graphite-based anode system. The analysis is conducted under open-circuit conditions, with the OCV determined from the SoC--OCV relation in the absence of an externally applied current. Prior to investigating the dense-to-porous transition, we validate the deterministic transport-limited growth predicted by the model against the analytical expression proposed by \citet{single_revealing_2017}. The noise map is then studied to avoid mesh-dependent perturbations and to control the spatial scale of the stochastic noise. The validated model is extended with the noise map to investigate the dense-to-porous transition. The transition is characterized using a roughness threshold based on the characteristic SEI molecular edge length. The roughness evolution is examined with respect to SEI thickness and time across different SoC conditions. Finally, the effect of dynamic SoC conditions resulting from irreversible capacity loss on the transition time and regime behavior is examined.
 
    % \clearpage % can remove if you don't want a new page 
    \section{Methodology}
\subsection{Model overview}
% --- Task 2.1: Define the system ---
% Goal:
% Describe the common system of battery, EC electrolyte, our assumption (diffusion of Li >>>>>> e)
We consider a two-dimensional SEI/electrolyte interfacial model to describe transport-limited SEI growth under open-circuit voltage (OCV) conditions. In this framework, neutral lithium radicals ($\ce{Li^0}$) diffuse through the pre-formed SEI layer toward the SEI/electrolyte interface, where they react with electrolyte species and contribute to further SEI formation. The phase-field modeling is used to track the evolving SEI/electrolyte interface, allowing the model to capture changes in SEI morphology during growth.

\begin{figure}[!ht]
    \centering
    \includegraphics[width=90 mm]{"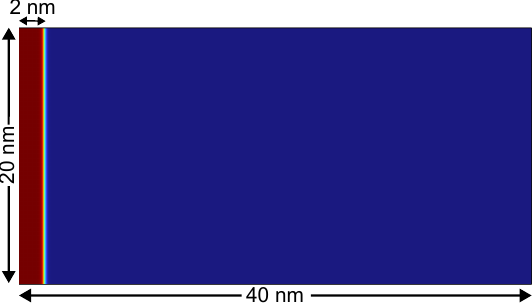"}
    \caption{Initial geometry of the simulation cell consisting of \SI{2}{nm} \ce{Li2EDC} SEI layer and \SI{38}{nm} EC electrolyte region.}
    \label{fig:system_introduction}
\end{figure}

We investigated an SEI/electrolyte interfacial system consisting of a pre-formed SEI layer, e.g., \ce{Li2EDC}, and an EC electrolyte. The initial thickness of the SEI was set to \SI{2}{nm} to prevent quantum tunneling effects \cite{soto_understanding_2018}. As shown in Figure~\ref{fig:system_introduction}, the two-dimensional simulation cell has dimensions of \SI{40}{nm} in the $x$ direction, corresponding to the SEI growth direction, and \SI{20}{nm} in the $y$ direction, corresponding to the cell height, with an initial \SI{2}{nm}-thick SEI layer defined along the $x$ direction and a \SI{38}{nm} electrolyte region.
Under open-circuit voltage (OCV) conditions, the external circuit is disconnected, meaning that electron transport is not driven by a macroscopic electric field. Instead, electrons are transported through the SEI bulk toward its surface. Here, without loss of generality, we assume that electrons appear in the form of neutral lithium radicals ($\ce{Li^0}$) \cite{soto_formation_2015, kobbing_growth_2023} driven entirely by a diffusion process. Previous studies indicate that the diffusion of these $\ce{Li^0}$ radicals is significantly slower than the transport of $\ce{Li+}$ ions \cite{von_kolzenberg_transition_2022, soto_understanding_2018, shi_defect_2013, shi_direct_2012}. Therefore, $\ce{Li^0}$ radical diffusion is treated as the transport-limited process in the present model and represents the effective transport of electric charge through the SEI. Accordingly, we assume that, in the present system, $\ce{Li+}$ ions are uniformly and sufficiently supplied to the SEI surface. In this study, SEI formation is represented by the following model reaction:
\begin{equation}
    \ce{2Li+ + 2e- + 2EC -> Li2EDC + C2H4}
    \label{eq:SEI_reaction}
\end{equation}

\subsection{Phase-field method}
% --- Task 2.2: phase-field method description ---
% Goal: phase-field method (Butler-volmer equation and model simplification (first order of taylor series)
We model SEI growth using a phase-field approach, in which an order parameter $\xi \in [0, 1]$ distinguishes the two phases of the system: $\xi = 0$ corresponds to the electrolyte and $\xi = 1$ to the \ce{Li2EDC} SEI. The interface between the two phases is represented as a diffuse region of finite thickness, where $\xi$ varies continuously between these limits. The evolution of $\xi$ is governed by the Butler-Volmer equation,
\begin{equation}
    I = -I_0 \left[\exp\left\{\frac{(1-\alpha)nF\eta}{R_\mathrm{gas}T}\right\}-\exp\left\{\frac{(-\alpha)nF\eta}{R_\mathrm{gas}T}\right\} \right].
    \label{eq:bv_advanced}
\end{equation}
The exchange current density is defined as $I_0=i_0\left(\frac{C_e}{C_{e,0}}\right)^{1-\alpha}$ depending on the electron concentration at the interface \cite{bazant_theory_2013}, where $i_0$ is the exchange current density in the bulk, $C_e$ is $\ce{Li^0}$ radicals concentration, $C_{e,0}$ is reference $\ce{Li^0}$ radicals concentration, $\alpha$ is the charge-transfer symmetry coefficient, and $n$ is the number of electrons transferred. The gas constant $R_\mathrm{gas}$ and Faraday constant $F$ are defined as \SI{8.314}{J/(mol\cdot K)} and \SI{96485}{C/mol}. The overpotential $\eta$ is defined as Eq.~\ref{eq:overpotential_definition}. By coupling the Butler--Volmer reaction kinetics to the phase-field formulation, the evolution of the order parameter \cite{liang_nonlinear_2014, chen_modulation_2015} is given by
\begin{equation}
    \frac{\partial \xi}{\partial t} = -I_0 \frac{V_m\gamma}{nF\kappa} h^\prime(\xi) \left[\exp\frac{(1-\alpha)nF\eta}{R_\mathrm{gas}T}-\exp\frac{(-\alpha)nF\eta}{R_\mathrm{gas}T} \right].
    \label{eq:bv_basic}
\end{equation}
The interfacial velocity $\frac{\partial\xi}{\partial t}$ is related to Butler-Volmer equation via molar volume of \ce{Li2EDC} $V_m$ and the gradient coefficient $\kappa=\frac{3}{2}\gamma \delta$, following \cite{liang_nonlinear_2014, jana_dendrite-separator_2015, cogswell_quantitative_2015, jeon_phase-field_2022}. Here, $\gamma$ is the surface energy of SEI, and \(\delta\) is the interfacial thickness. The kinetic term is weighted by $h^\prime(\xi) = 6\xi(1-\xi)$, which is the derivative of the interpolation function $h(\xi) = \xi^2(3-2\xi)$, which is a smooth interpolation between the bulk phases. Furthermore, $h^\prime(\xi)$ is a localized interpolation function of the order parameter that confines the reaction to the interface \cite{chen_modulation_2015, hong_phase-field_2018}. This ensures that the reaction vanishes in the bulk electrolyte and electrode and is non-zero only across the interface. To improve numerical stability and convergence, we simplify the equation using the first order of the Taylor series to have a stable numerical calculation and convergence
\begin{equation}
    \frac{\partial \xi}{\partial t} = -I_0\frac{V_m\gamma}{\kappa R_\mathrm{gas} T}h^\prime(\xi)\eta .
    \label{eq:first_taylor}
\end{equation}
The local overpotential $\eta$ extends the classic Butler-Volmer formulation to capture the coupled effects of interfacial and thermodynamic energy on the reaction kinetics

\begin{equation}
    \eta = -\left[h^\prime(\xi)\left(U_0+\frac{R_\mathrm{gas}T}{nF}\ln\tilde{C}_\mathrm{e}\right)-\frac{V_m}{nF}g^\prime (\xi) + \nabla \cdot \frac{V_m \kappa}{nF}\nabla \xi\right].
    \label{eq:overpotential_definition}
\end{equation}

The overpotential $\eta$ in Eq.~(\ref{eq:overpotential_definition}) captures three contributions to the reaction driving force. The first term reflects the local electrochemical driving force: $U_0 = \SI{0.8}{V}$ is the onset potential for SEI formation \cite{von_kolzenberg_transition_2022, kobbing_growth_2023, wang_review_2018, single_revealing_2017}, and $\ln\tilde{C}_\mathrm{e}$ accounts for the local $\ce{Li^0}$ radical concentration, where $\tilde{C}_\mathrm{e} = C_e / C_{e,0}$ is the activity of $\ce{Li^0}$ radicals. Despite the solid-state nature of the SEI, a dilute solution model is adopted,  assuming that interactions among mobile solute species are negligible. The reaction is localized to the diffuse interface via $h^\prime(\xi)$. The remaining two terms arise from the free energy of the phase-field: $g^\prime(\xi)$, where $g(\xi) = W\xi^2(1-\xi)^2$ is the double-well potential, and the gradient term $\nabla \cdot \kappa \nabla\xi$ acts together to separate the phases and maintain a stable interface.

\subsection{Diffusion model}
% --- Task 2.3: diffusion model coupling with reaction term ---
% Goal: describe the diffusion model involving the conservation law from reaction kinetics. / Boundary conditions
The evolution of the $\ce{Li^0}$ radicals concentration $C_e$ in the SEI is governed by a species conservation equation that couples Fickian diffusion, noise flux from $\ce{Li^0}$ radicals diffusion, and the consumption by the interfacial reaction. $\ce{Li^0}$ radicals transport in the SEI is modeled as Fickian diffusion with a stochastic flux contribution. First-principles studies \cite{wang_review_2018, soto_formation_2015} have shown that $\ce{Li^0}$ radical transport through \ce{Li2EDC} at the atomistic level proceeds via stochastic hopping between localized defect states rather than continuous electron transport. The noise term provides a continuum-level representation of this hopping mechanism, allowing the model to capture interface instability that pure transport without random noise cannot reproduce,

\begin{equation}
    \frac{\partial C_e}{\partial t} 
    = \nabla \cdot \left[D_e\,\nabla C_\mathrm{e} + \tilde{\lambda}(x,y)\right]
    -\,n\, C_{s}^\mathrm{SEI}\, h'(\xi)\, \frac{\partial \xi}{\partial t}
    \label{eq:species_transport}
\end{equation}
Here, $D_e$ and $\tilde{\lambda}(x,y)$ are the diffusivity of $\ce{Li^0}$ radicals in the SEI and the stochastic noise flux with an \SI{1e-11}{mol/m^2/s} amplitude, respectively, and $C_{s}^\mathrm{SEI}$ is the number density of molecules in the SEI (\ce{Li2EDC}). The last term in Eq.~\ref{eq:species_transport} represents the local consumption of $\ce{Li^0}$ radicals due to SEI formation at the interface, where $\frac{\partial \xi}{\partial t}$ is given by Eq.~(\ref{eq:first_taylor}). 

\subsection{Spatially random noise}
\begin{figure}[!ht]
    \centering
    \includegraphics[width=150 mm]{"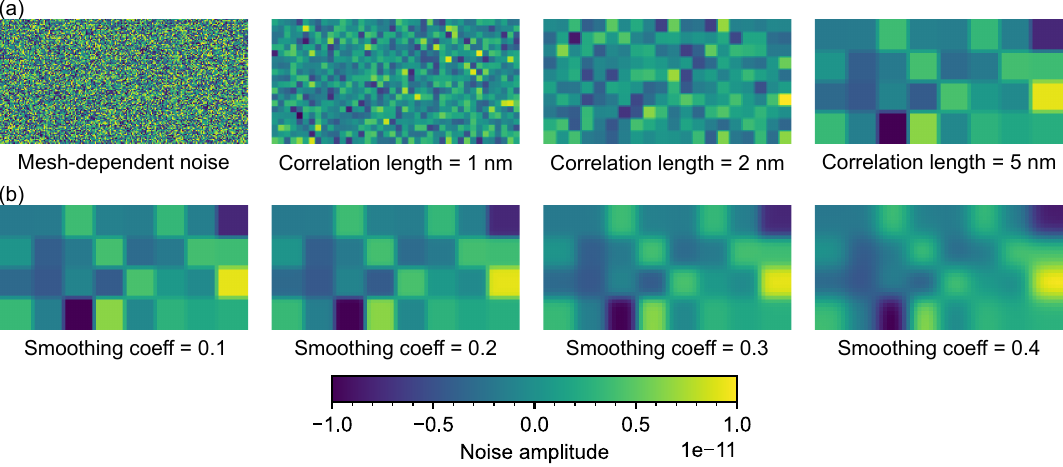"}
    \caption{Noise field generated for different correlation lengths $L_\mathrm{c}$ (a) and smoothing coefficients $s$ (b). The maximum amplitude is $A_\mathrm{max} = \SI{1e-11}{mol.m^{-2}.s^{-1}}$.}
    \label{fig:noise_input}
\end{figure}

Mesh-dependent random noise assigns an independent random value to each mesh node, as illustrated in the upper-left panel of Figure~\ref{fig:noise_input} (a). The spatial pattern of the noise depends on the mesh resolution because the random values are assigned node by node. To examine how the spatial correlation of the perturbation affects the interface roughness, a spatially correlated random field is generated over the domain. The spatial extent of the correlated perturbation is controlled by the correlation length, $L_\mathrm{c}$. For each correlation length, the domain is divided into square regions with a side length of $L_\mathrm{c}$ as illustrated in Figure~\ref{fig:noise_input} (a).
The noise lattice is extended by four lattice spacings beyond each boundary of the simulation domain, providing a sufficient buffer to minimize truncation of the Gaussian kernel near the domain edges. The continuous noise field $\tilde{\lambda}(\mathbf{x})$ is then derived using normalized Gaussian-kernel interpolation \cite{zhuang_statistical_2017},

\begin{equation} \tilde{\lambda}(\mathbf{x}) = \frac{\sum_{k} A_k \exp\!\left(-\dfrac{|\mathbf{x}-\mathbf{x}_k|^2} {2\sigma^2}\right)} {\sum_{k} \exp\!\left(-\dfrac{|\mathbf{x}-\mathbf{x}_k|^2} {2\sigma^2}\right)}, 
\end{equation} 
where \(A_k\) is the average of the mesh-dependent random noise within the \(k\)-th square region of side length \(L_\mathrm{c}\), and \(\mathbf{x}_k\) denotes the center position of that region. This construction maintains a similar spatial distribution of noise values despite variations in \(L_\mathrm{c}\), allowing the effect of correlation length on the interface roughness to be compared consistently. The smoothing factor is defined relative to the lattice spacing as $\sigma = s\,L_\mathrm{c}$, where the dimensionless smoothing coefficient $s$ controls how strongly neighboring noise lattices are blended. A small $s$ leaves each interpolated point dominated by its nearest noise lattice and preserves a block-like structure, whereas a larger $s$ blends multiple noise lattices and produces a smoother field with reduced local amplitude, as shown in Figure~\ref{fig:noise_input} (b). The field is finally rescaled so that its maximum absolute amplitude equals a prescribed value $A_\mathrm{max}$, which fixes the perturbation strength independently of $L_\mathrm{c}$ and $s$. The equation of noise field, $\lambda(\mathbf{x})$ is finally derived as follows:

\begin{equation} 
\lambda(\mathbf{x}) = A_\mathrm{max} \, 
\frac{\tilde{\lambda}(\mathbf{x})}{\max_{\mathbf{x}} |\tilde{\lambda}(\mathbf{x})|}.
\end{equation} 

% Figure~\ref{fig:noise_input} shows the resulting noise fields for a range of $L_\mathrm{c}$ (a) and $s$ (b). The same realization, generated with a fixed random seed, is used for all SoC cases so that the SoC-dependent response can be compared under an identical noise field.

\subsection{Capacity Loss \& Dynamic SoC}
The dynamic SoC is formulated by tracking the irreversible capacity loss during SEI growth \cite{single_identifying_2018}. Time-dependent SoC is quantified relative to the maximum capacity $Q_\mathrm{max}$ as
\begin{equation}
    \mathrm{SoC}(t) = \mathrm{SoC}(t_{n-1}) - 
    100\frac{Q_\mathrm{irr}(t)-Q_\mathrm{irr}(t_{n-1})}{Q_\mathrm{max}},
    \label{eq:dynamic_SOC}
\end{equation}
where $\mathrm{SoC}(t)$ is dynamic SoC, $t_{n-1}$ denotes the previous time step, and $Q_\mathrm{irr}(t)$ is the irreversible capacity loss. Following \citet{single_identifying_2018} and \citet{von_kolzenberg_transition_2022}, the total irreversible capacity is decomposed into an SEI-related contribution and a SEI-independent linear contribution:
\begin{equation}
    Q_\mathrm{irr}(t) = Q^{\mathrm{SEI}}_{\mathrm{irr}}(t) + Q^{\mathrm{lin}}_{\mathrm{irr}}(t)
    \label{eq:irreversible_volt}
\end{equation}
Here, \(Q^{\mathrm{SEI}}_{\mathrm{irr}}\) represents the irreversible lithium consumption associated with SEI formation, whereas \(Q^{\mathrm{lin}}_{\mathrm{irr}}\) accounts for the linear capacity loss that is assumed to be independent of the open-circuit voltage.
\begin{equation}
    Q^{\mathrm{SEI}}_{\mathrm{irr}}(t) = \sqrt{\frac{2sF^2}{V^2}} \cdot A_{\mathrm{el}} L(t) 
    \label{eq:Qirr_SEI_definition}
\end{equation}
Here, $L(t)$ follows the analytical SEI thickness derived by \citet{single_identifying_2018}:
\begin{equation}
    L(t)=\sqrt{V C_{e,0} D_e \exp\left[\frac{-e\cdot\mathrm{OCV_0}}{k_B T}\right]\cdot t + L_0^2}
    \label{eq:Length_theory}
\end{equation}
Here, \(A_{\mathrm{el}}\) is the active electrode surface area, and $V = N_\mathrm{A} a^3$ is the mean partial molar volume of SEI, where $a$ is the characteristic molecular edge length of the SEI species, and $s$ is the mean stoichiometric coefficient of $\ce{Li^0}$ radicals in the SEI formation reaction. In this analytical expression, $L(t)$ is evaluated using the initial open-circuit voltage (OCV$_0$), so that the voltage-dependent growth prefactor is fixed for each initial SoC condition. Considering the dynamic SoC, the resulting irreversible capacity loss then updates the SoC as a function of time, thereby incorporating SoC evolution into the dynamic-SoC simulations.

The linear capacity loss 
\begin{equation}
    Q^{\mathrm{lin}}_{\mathrm{irr}}(t) = \tau \cdot t
    \label{eq:Qlin_SEI_definition}
\end{equation}
reaches \SI{4.5}{\%} of the maximum capacity after 9.5 months, independently of the SoC \cite{single_identifying_2018}. This gives a linear capacity loss rate of $\tau = \SI{1.974e-5}{C/s}$.

\subsection{Calculation of the surface roughness}
The transition of the porous and dense SEI region is quantified by an interface roughness measure, defined as the root-mean-square (RMS) deviation of the interface position from its mean. In the phase-field framework, the interface is not a sharp boundary but the order parameter $\xi(x,y)$ transitions continuously between the bulk phases. We therefore define the interface position as the location where $\xi = 0.5$, corresponding to the midpoint between the two bulk values
\begin{equation}
    \Gamma = \{(x,y)\ |\ \xi(x,y) = 0.5\}.
    \label{eq:interface_definition}
\end{equation}
The interface is then sampled along the $y$-direction at $N$ locations. At each sampling line $y_i$, the distance from the SEI boundary to $\Gamma$ is measured as the local interface position. The local interface position $x_\Gamma(y_i)$ is obtained by linearly interpolating $\xi$ within the triangular element crossed by $\Gamma$, rather than restricting the position to mesh nodes. The same procedure is applied to all $N$ sampling lines, and the mean interface position is computed as
\begin{equation}
    x_{avg} = \frac{1}{N}\sum_{i=1}^{N} x_\Gamma(y_i).
\end{equation}
The roughness is then defined as the root-mean-square deviation of the individual sample positions
\begin{equation}
    R = \sqrt{\frac{1}{N}\sum_{i=1}^{N}\left[x_{\Gamma} (y_i)
    - x_{avg}\right]^2}.
    \label{eq:roughness_definition}
\end{equation}

\begin{figure}[!ht]
    \centering
    \includegraphics[width=90 mm]{"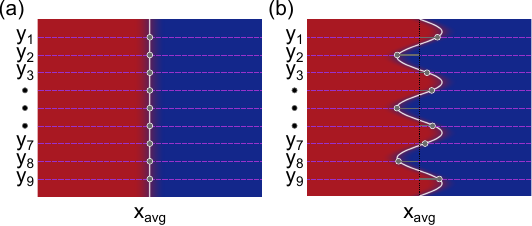"}
    \caption{Methodology for roughness assessment via multi-node sampling. (a) Flat interface; (b) curved interface.}
    \label{fig:roughness_evaluation}
\end{figure}

Figure~\ref{fig:roughness_evaluation} (a) and (b) illustrate the sampling procedure for nine representative samples $y_n$. In Figure~\ref{fig:roughness_evaluation} (a), the interface is planar, so all $x_\Gamma(y_i)$ coincide with $x_{avg}$ and roughness $R = 0$. In Figure~\ref{fig:roughness_evaluation} (b), the interface exhibits a wavy profile, the individual sample positions deviate from the mean interface position, and $R$ increases accordingly. The schematic uses nine samples for clarity; the actual evaluation employs a substantially larger $N$, with 240 sample points in this assessment, to ensure converged statistics. This roughness measure allows us to quantify the morphological state of the SEI/electrolyte interface. It is used to identify the onset of the instability regime in which SEI formation becomes a porous region.

% --- Task 2.5: Introduce the assessment ---
% Goal: 
% 1. Describe how to get the roughness indicator
% 2. Describe the method to measure the dynamic state of charge

\subsection{Numerical setting and simulation parameters}
% --- Task 2.4: computational model parameter ---
% Goal: Introduce the model parameters involving the OCV-SOC plot

\begin{figure}[!ht]
    \centering
    \includegraphics[width=90 mm]{"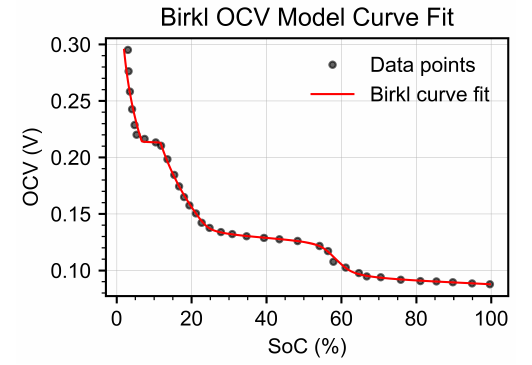"}
    \caption{Open-circuit voltage (OCV) as a function of the battery state of charge (SOC). The data points are obtained from \citet{von_kolzenberg_transition_2022}, and the solid line represents the approximation using the Birkl model \cite{birkl_parametric_2015}.}
    \label{fig:OCV_iapproximation}
\end{figure}

% --- Task 2.6: Introduce the assessment ---
% Goal: Describe how to get the roughness indicator
The phase-field model described thus far represents SEI growth in terms of the electrochemical driving force determined by the OCV. However, under open-circuit conditions, the OCV is not an independent input parameter but is determined by the battery state of charge (SoC). Therefore, incorporating the SoC--OCV relationship into the model enables the analysis of how the $\ce{Li^0}$ radical supply, SEI growth rate, and interfacial instability vary under different SoC conditions.
The SoC--OCV relationship of the graphite anode is obtained by fitting the parametric model of \citet{birkl_parametric_2015} to SoC--OCV data points extracted from \citet{von_kolzenberg_transition_2022}. The fitted function returns SoC as a function of OCV, following the explicit form given in Eq.~\eqref{eq:soc-ocv}. This relationship must be inverted to provide $\mathrm{OCV}(\mathrm{SoC})$ for use in the simulation. The inversion is performed numerically: 101 uniformly spaced SoC values are sampled in the range \SI{0}{\%} -- \SI{100}{\%}, and the corresponding OCV is obtained by solving Eq.~\eqref{eq:soc-ocv} at each sample point. The resulting pairs are stored as a lookup table, and for SoC values between sampled points, OCV is obtained by linear interpolation between the data points. The fitted curve is compared with the original data points in Figure~\ref{fig:OCV_iapproximation}, confirming good agreement across the full SoC range.

\begin{equation}
\begin{split}
    \mathrm{SoC}(\mathrm{OCV}) = &\ \frac{100.00}{1 + \exp\!\left[\dfrac{(\mathrm{OCV} - 0.0457) \cdot 0.4011\, e}{k_B T}\right]} \\
    &+ \frac{53.66}{1 + \exp\!\left[\dfrac{(\mathrm{OCV} - 0.0890) \cdot 11.0391\, e}{k_B T}\right]} \\
    &+ \frac{26.83}{1 + \exp\!\left[\dfrac{(\mathrm{OCV} - 0.1288) \cdot 11.6884\, e}{k_B T}\right]} \\
    &+ \frac{4.86}{1 + \exp\!\left[\dfrac{(\mathrm{OCV} - 0.2136) \cdot 198.1758\, e}{k_B T}\right]}.
\end{split}
\label{eq:soc-ocv}
\end{equation}

Next, Table~\ref{tab:parameters} shows the definition of the parameters and the units.
\begin{table}[!ht]
    \centering
    \caption{Constant parameters of the model}
    \label{tab:parameters}
    \begin{tabular}{l l l l l}
        \hline
        Symbol & Description & Value & Unit & Source \\
        \hline
        $F$         & Faraday constant                               & \SI{96485}{}           & \SI{}{C/mol}      & --- \\
        $R_\mathrm{gas}$         & Gas constant                                   & \SI{8.314}{}           & \SI{}{J/(mol.K)}  & --- \\
        $k_B$       & Boltzmann constant                             & \SI{1.381e-23}{}       & \SI{}{J/K}        & --- \\
        $N_\mathrm{A}$ & Avogadro number                             & \SI{6.022e+23}{}       & \SI{}{1/mol}      & --- \\
        $e$         & Elementary charge                              & \SI{1.602e-19}{}       & \SI{}{C}          & --- \\
        $T$         & Temperature                                    & \SI{298.15}{}             & \SI{}{K}          & --- \\
        \hline
        $V_m$       & Molar volume of \ce{Li2EDC}                    & \SI{9.62e-5}{}         & \SI{}{m^3/mol}    & \cite{cipolla_effect_2022} \\
        $a$         & Characteristic edge length of \ce{Li2EDC}      & \SI{5.4e-10}{}         & \SI{}{m}          & \cite{keil_calendar_2016} \\
        $V$         & Mean partial molar volume ($= N_\mathrm{A} a^3$) & \SI{9.47e-5}{}       & \SI{}{m^3/mol}    & derived \\
        $C_s^\mathrm{SEI}$ & SEI molar concentration                  & \SI{10395}{}           & \SI{}{mol/m^3}    & derived \\
        $n$         & Electrons per \ce{Li2EDC} unit                 & \SI{2}{}               & ---               & stoichiometry \\
        $s$         & Mean stoichiometric coeff. of $e^-$            & \SI{2}{}               & ---               & stoichiometry \\
        $U_0$       & EC reduction onset potential                   & \SI{0.8}{}             & \SI{}{V}          & \cite{von_kolzenberg_transition_2022, kobbing_growth_2023} \\
        \hline
        $D_e$       & Effective $\ce{Li^0}$ radicals diffusivity         & \SI{1e-15}{}           & \SI{}{m^2/s}      & \cite{von_kolzenberg_transition_2022} \\
        $C_{e,0}$ & Reference $\ce{Li^0}$ radicals concentration & \SI{0.01}{}           & \SI{}{mol/m^3}    & \cite{von_kolzenberg_transition_2022} \\
        $i_0$       & Exchange current density in bulk               & \SI{4e-5}{}            & \SI{}{A/m^2}      & assumption \\
        $\alpha$    & Charge transfer symmetric coefficient                    & \SI{0.5}{}             & ---               & symmetric \\
        \hline
        $\delta$    & Interfacial thickness                          & \SI{8e-10}{}           & \SI{}{m}          & --- \\
        $W$         & Double-well barrier height ($=8\kappa/\delta^2$) & \SI{7.5e9}{}          & \SI{}{J/m^3}     & derived \\
        $\kappa$    & Gradient energy coefficient ($= \frac{3}{2}\gamma \delta$) & \SI{6e-10}{} & \SI{}{J/m}      & derived \\
        $\gamma$    & Surface energy of SEI                          & \SI{0.5}{}             & \SI{}{J/m^2}      & assumption \\
        \hline
        $A_\mathrm{el}$ & Active electrode surface area              & \SI{14.34}{}           & \SI{}{m^2}        & \cite{von_kolzenberg_transition_2022} \\
        $Q_\mathrm{max}$ & Maximum cell capacity                     & \SI{10800}{}           & \SI{}{C}          & \cite{von_kolzenberg_transition_2022} \\
        $\tau$      & Linear capacity loss rate                      & \SI{1.974e-5}{}        & \SI{}{C/s}        & \cite{single_identifying_2018} \\
        \hline
    \end{tabular}
    \label{tab:parameters}
\end{table}
We note that the exchange current density $i_0$ is treated as an assumed parameter because direct experimental constraints on the $\ce{Li^0}$ radical reduction reaction at the SEI/electrolyte interface remain limited. This treatment is justified by the transport-limited SEI growth regime reported in previous studies \cite{single_revealing_2017, single_identifying_2018, von_kolzenberg_transition_2022}, where reactant transport through the SEI primarily governs the growth rate rather than interfacial reaction kinetics. Accordingly, the simulation results are insensitive to the specific value of $i_0$ within the kinetically relevant range considered here. Similarly, the surface energy of \ce{Li2EDC} is treated as an effective SEI interfacial parameter rather than as a material property. In this work, \(\gamma_{\mathrm{SEI}}=\SI{0.5}{J.m^{-2}}\) is used as the assumption.

The SEI formation model using phase-field solves the two domains: the order parameter and the $\ce{Li^0}$ radical concentration. In addition to the governing equations specified in Eqs.~\eqref{eq:first_taylor} and \eqref{eq:species_transport}, appropriate initial and boundary conditions must be established for each of these parameters. The boundary conditions of the two domains are shown in Table~\ref{tab:boundary_conditions}.

\begin{table}[!ht]
    \centering
    \caption{Summary of boundary conditions for the model}
    \label{tab:boundary_conditions}
    \begin{tabular}{l l l}
        \hline
        Parameter & Boundary ($nm$) & Boundary Expression \\
        \hline
        $\xi$     & $x = 0$                 & $\xi = 1$ \\
                  & $x = 40$                & $\xi = 0$ \\
                  & $y = 0$                 & $\nabla\xi\cdot \hat{u} = 0$ \\
                  & $y = 20$                & $\nabla\xi\cdot \hat{u} = 0$ \\
        \hline
        $C_e$     & $x = 0$                 & $C_e = C_{e,0} \exp\left[\frac{-F\cdot\mathrm{OCV(\mathrm{SoC})}}{R_\mathrm{gas}T} \right]$ \\
                  & $x = 40$                & $\nabla C_e \cdot \hat{u} = 0$ \\
                  & $y = 0$                 & $\nabla C_e \cdot \hat{u} = 0$ \\
                  & $y = 20$                & $\nabla C_e \cdot \hat{u} = 0$ \\
        \hline
    \end{tabular}
\end{table}
The boundary conditions for $\xi$ are 1 at the SEI phase and 0 at the electrolyte phase. The concentration boundary condition for $C_e$ at $x = 0$ follows from the Nernst equation, relating the equilibrium $\ce{Li^0}$ radicals concentration at the electrode/SEI interface to the open-circuit voltage $\mathrm{OCV}(\mathrm{SoC})$\cite{von_kolzenberg_transition_2022, single_revealing_2017}. At the outer boundary ($x = \SI{40}{nm}$), a no-flux condition is imposed for $\ce{Li^0}$ radicals concentration, ensuring that $\ce{Li^0}$ radicals entering the domain at $x = 0$ are consumed by SEI formation reactions within the domain rather than lost to the bulk electrolyte phase, consistent with the formulation of \citet{single_revealing_2017}. Symmetric no-flux conditions are applied at $y = 0$ and $y = 20$~nm. 

The initial $\xi$ shown in Figure~\ref{fig:system_introduction} is represented by
\begin{equation}
 \xi(x,y)=\frac{\tanh{\left[-6\cdot (x-\SI{2}{nm})/\mathrm{nm}\right]}}{2}+0.5.
 \label{eq: initial_order}
\end{equation}
The initial $C_e$ is defined as
\begin{equation}
    C_e(x,y) = C_{e,0} \exp\left[\frac{-F\cdot\mathrm{OCV(\mathrm{SoC})}}{R_\mathrm{gas}T} \right] \cdot \left[\frac{\tanh{\left\{-6\cdot (x-\SI{2}{nm})/\mathrm{nm}\right\}}}{2}+0.5\right].
\label{eq:initial_ce}
\end{equation}
Additionally, we used a uniform quad mesh element of \SI{0.2}{nm} by \SI{0.2}{nm} for the order parameter and concentration domains. 
    \section{Results \& Discussion}
We investigate the morphological evolution of the SEI governed by the transport-limited diffusion of $\ce{Li^0}$ radical. First, we validate the phase-field framework by comparing the time-dependent SEI thickness against the analytical growth model of \citet{single_identifying_2018} (Eq.~\ref{eq:Length_theory}). Next, we analyze the effect of the spatial random noise map using the correlation length and the smoothing coefficient. A stochastic noise map based on the noise investigation is applied to the diffusion flux to trigger interfacial instability across states of charge (SoC) ranging from \SI{20}{\%} to \SI{80}{\%} in \SI{5}{\%} increments. By tracking the interface roughness, we evaluate the transition time from dense to porous SEI morphology using multiple roughness thresholds. Finally, we incorporate time-dependent capacity loss under open-circuit conditions \cite{single_identifying_2018} to evaluate its influence on the transition to porous SEI.

\subsection{Validation of SEI growth}
We first validate our model reproducing the analytical SEI growth behavior reported by \citet{single_identifying_2018}. The analytical growth law of \citet{single_identifying_2018} (Eq.~\ref{eq:Length_theory}) describes deterministic, transport-limited SEI growth. Therefore, the stochastic noise term is deactivated for validation. The SEI/electrolyte interface position is defined as the location where the interpolation-function derivative $h'(\xi)$ reaches its maximum; for the interpolation function used in this study, this maximum occurs at $\xi = 0.5$.

\begin{figure}[!ht]
    \centering
    \includegraphics[width=90 mm]{"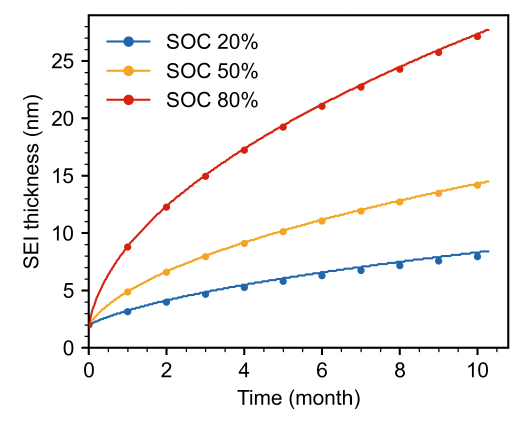"}
    \caption{Time-dependent SEI layer thickness at three states of charge (SoC). The circular symbols represent the analytical growth law evaluated from \citet{single_identifying_2018}, while the solid lines denote the results obtained from the present phase-field model.}
    \label{fig:SEI_thickness_rate}
\end{figure}

Figure~\ref{fig:SEI_thickness_rate} compares the present simulation results with the analytical growth law for three representative states of charge (SoC \SI{20}{\%}, \SI{50}{\%}, and \SI{80}{\%}) over approximately ten months. The analytical reference was evaluated at one-month intervals based on Eq.~\eqref{eq:Length_theory}. The good agreement between the phase-field results and the analytical model in Figure~\ref{fig:SEI_thickness_rate} shows that the deterministic simulation reproduces the characteristic $\sqrt{t}$ dependence of transport-limited SEI growth. This agreement also indicates that the boundary conditions for $\ce{Li^0}$ radical concentration determined from the Nernst relation and the SoC--OCV relation (Eq.~\ref{eq:soc-ocv}) used in the model are consistent with the analytical reference within the investigated SoC range. In addition, the result suggests that the first-order Taylor approximation of the Butler--Volmer equation is adequate for reproducing the deterministic transport-limited SEI growth considered here.

By reproducing the analytical transport-limited growth behavior, this deterministic validation establishes a consistent baseline for the effective transport properties ($D_e$, $C_{e,0}$) and interfacial reaction parameters ($\gamma$, $i_0$) adopted in the present phase-field model. Based on this baseline, the following subsection introduces stochastic noise into the diffusion flux to analyze the onset of interfacial instability and the resulting transition from dense to porous SEI morphology.

\subsection{Sensitivity of the dense-to-porous transition to the noise field parameters}
\begin{figure}[!ht]
    \centering
    \includegraphics[width=90 mm]{"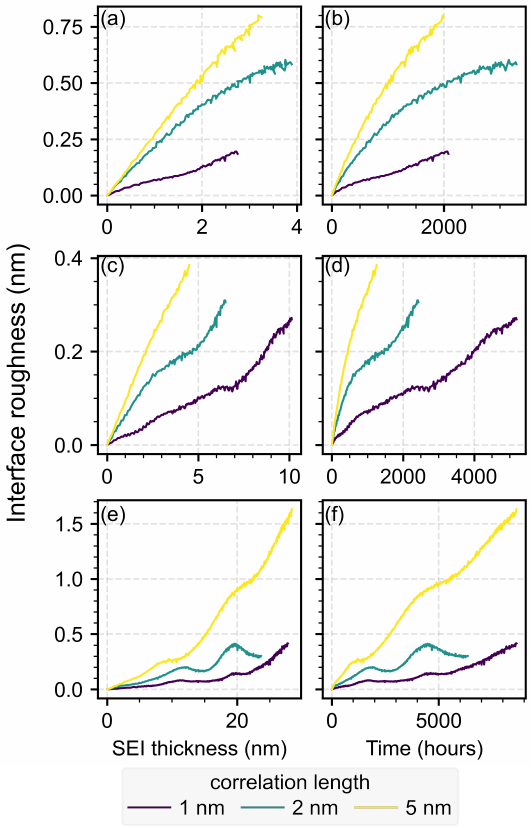"}
    \caption{Effect of the noise correlation length on the interface roughness at three states of charge. Each row shows the roughness as a function of SEI thickness (left) and time (right) for (a,~b)~SoC \SI{20}{\%}, (c,~d)~SoC \SI{50}{\%}, and (e,~f)~SoC \SI{80}{\%}. The maximum noise amplitude and the smoothing coefficient are identical in all cases as \SI{1e-11}{\mole/(\meter^2\second)} and \SI{0.4}{}.}
    \label{fig:noise_effect}
\end{figure}

Taking spatially random noise into account, Figure~\ref{fig:noise_effect} shows that the interface roughness generally increases as the SEI grows, both with increasing SEI thickness and with time. Although local fluctuations and temporary reductions in roughness are observed under some conditions, the overall evolution is characterized by progressive roughening of the SEI/electrolyte interface. The magnitude and pattern of this roughness evolution depend on the noise correlation length and SoC.

The sensitivity of the interface evolution to the spatial correlation of the noise field was evaluated before investigating the SoC effects. To examine the effect of the spatial correlation of the noise, the interface roughness was compared for correlation lengths of $L_\mathrm{c}=1$, $2$, and \SI{5}{\nano\meter} across different SoC values. In all cases, the maximum noise amplitude was fixed at $A_\mathrm{max}=\SI{1e-11}{\mole/(\meter^2\second)}$. For all SoC values, the interface roughness increased with higher $L_\mathrm{c}$. By contrast, the smoothing coefficient showed no clear effect on the overall roughness trend, as presented in the Supplementary Information (Figure~\ref{fig:noise_effect_smoothing_factor}).

Increasing $L_\mathrm{c}$ consistently increases the roughness amplitude, whereas its effect on the roughness pattern varies with SoC. At SoC \SI{20}{\%} and \SI{50}{\%} in Figure~\ref{fig:noise_effect} (a-d), there are no distinct local maxima or minima observed for any range of the correlation lengths. The same behavior is observed at SoC \SI{80}{\%} for $L_\mathrm{c}=\SI{5}{\nano\meter}$ in Figure~\ref{fig:noise_effect} (e-f). In contrast, at SoC \SI{80}{\%}, local maxima appear at SEI thicknesses of approximately \SI{10}{\nano\meter} and \SI{20}{\nano\meter} for $L_\mathrm{c}=\SI{1}{\nano\meter}$ and \SI{2}{\nano\meter}, respectively. These local maxima reflect a transient competition between surface relaxation and noise-induced roughening. As the correlation length increases, the perturbation extends over wider spatial regions, so the interfacial valleys become too broad to be filled by the \ce{Li^0} radical. Consequently, noise-induced roughening becomes dominant over surface relaxation, suppressing the development of distinct local maxima. This behavior is observed at lower SoC values and at SoC \SI{80}{\%} for $L_\mathrm{c}=\SI{5}{\nano\meter}$.

According to \citet{von_kolzenberg_transition_2022, keil_calendar_2016}, the characteristic edge length of an SEI molecule of \ce{Li2EDC} is approximately \SI{5.42}{\angstrom}. In this study, the interface roughness is defined as the root-mean-square deviation of the SEI height from its mean value. The molecular edge length therefore provides a physically motivated reference scale for evaluating whether the interfacial deviation becomes comparable to the characteristic size of the SEI species. Half of the molecular edge length, \SI{2.71}{\angstrom}, was adopted as the threshold for identifying the onset of the porous regime in our analysis. 
At the same time, the SoC-dependent competition between surface relaxation and noise-induced roughening remains observable when the correlation length is sufficiently small, whereas larger $L_\mathrm{c}$ increasingly suppresses this competition. This indicates that an excessively large correlation length is not suitable for the noise map implementation. Among the tested noise maps, a correlation length of $L_\mathrm{c}=\SI{2}{\nano\meter}$ resembles the typical size of grains inside the SEI and consistently reaches approximately \SI{0.3}{\nano\meter} of interface roughness while retaining the SoC-dependent features of the roughness evolution. Therefore, $L_\mathrm{c}=\SI{2}{\nano\meter}$ was selected for the noise map used in the subsequent static- and dynamic-SoC simulations.

\subsection{Evolution of interface roughness and phase transition}

\begin{figure}[!ht]
    \centering
    \includegraphics[width=150 mm]{"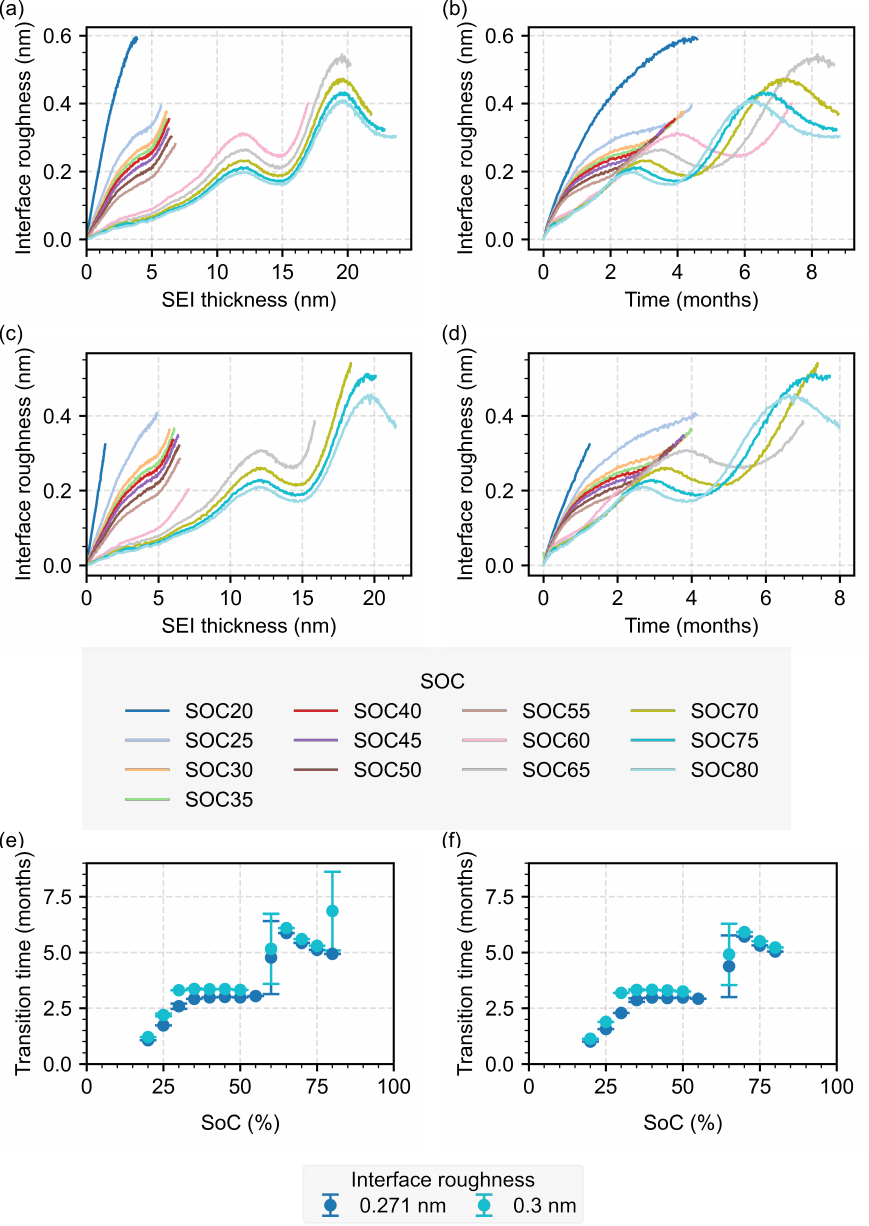"}
    \caption{Comparison of interface roughness evolution and dense-to-porous transition time under static and dynamic SoC conditions. (a, c) Interface roughness according to the SEI thickness under (a) static and (c) dynamic SoC. (b, d) Temporal evolution of the interface roughness $R$ under (b) static and (d) dynamic SoC conditions for SoC values ranging from \SI{20}{\%} to \SI{80}{\%} in \SI{5}{\%} increments. In the dynamic case, SoC denotes the initial SoC, $\mathrm{SoC}_0$, and decreases over time due to irreversible capacity loss. (e, f) Dense-to-porous transition time under (e) static and (f) dynamic SoC conditions, extracted from the roughness curves in (b) and (d), respectively, using two threshold values, $R_\mathrm{th} = $ \SI{0.271}{\nano\meter} and \SI{0.3}{\nano\meter}. For each threshold, the error bars indicate the time interval between the first and last crossings of the threshold value. The dynamic SoC result for $\mathrm{SoC}_0 = \SI{55}{\%}$ and $\mathrm{SoC}_0 = \SI{60}{\%}$ at $R_\mathrm{th} =$ \SI{0.3}{\nano\meter} and the static Soc result for \SI{60}{\%} at $R_\mathrm{th} =$ \SI{0.271}{\nano\meter} are unavailable because $R$ did not reach the threshold within the simulation period.}
    \label{fig:transition_time}
\end{figure}

The effect of SoC on the dense-to-porous SEI transition was investigated using stochastic noise with a correlation length of \SI{2}{\nano\meter} and an amplitude of \SI{1e-11}{\mole/(\meter^2\second)}, following the validation of deterministic SEI growth in Section~3.1. SEI formation simulations were performed for SoC values ranging from \SI{20}{\%} to \SI{80}{\%} in \SI{5}{\%} increments. The temporal evolution of the interface roughness $R$ (Eq.~\eqref{eq:roughness_definition}) was then analyzed as a quantitative indicator of the dense-to-porous transition.

Figure~\ref{fig:transition_time} (a) reveals a common roughness pattern across different SoC conditions. Fluctuations in the interface roughness are observed at similar SEI thicknesses across the SoC cases, while the overall roughness evolution exhibits distinct patterns within three SoC regimes. In regime~\rom{1} at SoC \SI{20}{\%}, $R$ enters a sustained roughness-growth trend from the beginning of the simulation. In regime~\rom{2}, corresponding to SoC \SI{25}{\%}--\SI{55}{\%}, a fluctuation appears near an SEI thickness of approximately \SI{5}{\nano\meter}, followed by a locally reduced gradient of $R$. In regime~\rom{3}, noise-induced perturbations remain relatively suppressed at \SI{5}{\nano\meter}, but they become amplified at a later growth stage near \SI{11}{\nano\meter}, \SI{15}{\nano\meter}, and \SI{19}{\nano\meter}. The occurrence of fluctuations at similar SEI thicknesses is associated with applying the same spatial noise pattern to the diffusion flux for all SoC cases. The imposed noise locally increases or decreases the $\ce{Li^0}$ radical flux, leading to surface instability. Sustained supply of $\ce{Li^0}$ radicals can reduce the SEI thickness variation and promote surface relaxation. Here, surface relaxation refers to this reduction of the SEI thickness variation at the interface. These trends reflect a competition between noise-induced transport and interfacial reaction kinetics, characterized by a balance between noise-induced roughening and surface relaxation. The variation of this balance across the SoC range is associated with the SoC-dependent supply of $\ce{Li^0}$ radicals to the interface and broadly corresponds to the variation of OCV in the graphite SoC--OCV relation in Figure~\ref{fig:OCV_iapproximation}. Figure~\ref{fig:transition_time}(b) shows the temporal evolution of the interface roughness $R$ under static SoC conditions. Unlike the thickness-based representation in Figure~\ref{fig:transition_time}(a), the time-dependent curves reflect both morphological evolution and the SoC-dependent SEI growth rate. The tipping-point features that are separated by SoC in Figure~\ref{fig:transition_time}(a) become less aligned and partly overlap in the time domain, particularly in regime~\rom{3}. This trend in the time domain arises from the $\sqrt{t}$ dependence of the diffusion-controlled SEI growth described by Eq.~\ref{eq:Length_theory}.

To quantify the dense-to-porous transition time, two roughness thresholds were introduced: $R_\mathrm{th} = $ \SI{0.271}{\nano\meter} and \SI{0.3}{\nano\meter}. The transition time generally increases with SoC, from approximately one month at SoC \SI{20}{\%} to more than seven months at SoC \SI{80}{\%}; however, this increase is not uniform across the investigated SoC range. The most pronounced step change in the transition time appears in the SoC \SI{55}{\%}--\SI{60}{\%} range in Figure~\ref{fig:transition_time}(e), where the system changes from regime~\rom{2} to regime~\rom{3}. This transitional SoC range is also reflected in the Birkl SoC--OCV curve, where the OCV changes from approximately \SI{0.120}{\volt} to \SI{0.102}{\volt} over the SoC \SI{55}{\%}--\SI{60}{\%} range. Through the Nernst boundary condition, the decrease in OCV increases the boundary $\ce{Li^0}$ radical concentration by approximately a factor of 2.05. The resulting increase in \ce{Li^0} radical supply accelerates SEI growth while sustaining stronger surface relaxation, thereby delaying the time for the interface roughness to reach the dense-to-porous transition threshold. This direct correspondence indicates that the dense-to-porous transition time is closely linked to the SoC--OCV relation. Since the SoC--OCV relation depends on the electrode chemistry, cell configuration, and battery specifications, the SoC-dependent trend of the dense-to-porous transition time is also expected to vary across battery systems. Therefore, the critical SoC range identified here should be interpreted as specific to the graphite-based system considered in this study. Nevertheless, the same analysis framework can be extended to other systems by using the corresponding SoC--OCV relation and recalibrating the relevant transport and reaction parameters. 

Overall, these results indicate that the dense-to-porous transition is not a single boundary defined by one roughness threshold, but an SoC-dependent transition behavior. The correspondence between the transition pattern and the SoC--OCV relation further shows that this relation is essential for interpreting SEI states. However, the present analysis is based on a static SoC assumption, whereas the SoC can decrease over time under open-circuit storage due to capacity loss. Therefore, the following subsection examines how capacity-loss-induced dynamic SoC conditions modify the dense-to-porous transition time and the associated regime boundaries.

\subsection{Effect of dynamic SoC degradation on the dense-to-porous transition}

To account for the time-dependent change in SoC during open-circuit storage, the SoC is allowed to decrease over time following the capacity loss model reported by \citet{single_identifying_2018} and \citet{von_kolzenberg_transition_2022}. The resulting SoC decrease modifies the effective OCV, which, through the Nernst boundary condition, determines the interfacial $\ce{Li^0}$ radical concentration and consequently affects the SEI growth rate. This dynamic SoC condition is used to evaluate how time-dependent SoC decrease affects the dense-to-porous transition pattern of the SEI.

Figure~\ref{fig:transition_time}(c) and (d) show the interface roughness  $R$ as a function of SEI thickness and time, respectively, under dynamic SoC conditions. Figure~\ref{fig:transition_time}(f) presents the corresponding dense-to-porous transition time. Here, SoC denotes the initial state of charge, $\mathrm{SoC}_0$, because the SoC decreases over time under dynamic conditions. Overall, the roughness evolution under dynamic SoC closely resembles that under static SoC, and the regime classification remains valid for most cases. Two exceptions appear at $\mathrm{SoC}_0 = \SI{25}{\%}$ and $\SI{60}{\%}$ in Figure~\ref{fig:transition_time}(c). At $\mathrm{SoC}_0 = \SI{25}{\%}$, the interface roughness increases nearly linearly, without the fluctuation observed under the corresponding static SoC condition in Figure~\ref{fig:transition_time}(a).
At $\mathrm{SoC}_0 = \SI{60}{\%}$, the roughness initially follows the regime~\rom{3} behavior, with the fluctuation near \SI{5}{\nano\meter} suppressed, but $R$ rises abruptly once the SEI reaches approximately \SI{6}{\nano\meter}. Figure~\ref{fig:transition_time}(d) further shows that the evolution of the interface roughness at $\mathrm{SoC}_0 = \SI{65}{\%}$ resembles that of the static SoC \SI{60}{\%} case in Figure~\ref{fig:transition_time}(b). This suggests that the transitional boundary between regime~\rom{2} and regime~\rom{3}, located around SoC \SI{60}{\%} under static conditions, shifts to a higher initial SoC under dynamic conditions.

Figure~\ref{fig:transition_time}(e) and (f) also compare the dense-to-porous transition times under static and dynamic SoC conditions for the two roughness thresholds. The influence of dynamic SoC varies across the investigated SoC range. In the low- and intermediate-SoC range, SoC \SI{20}{\%}--\SI{55}{\%}, the static and dynamic results are nearly identical because the accumulated capacity loss produces only a minor SoC change during the simulation period. In contrast, the most pronounced dynamic effect appears at SoC \SI{65}{\%}, where the dynamic SoC condition advances the transition by approximately \SI{1}{month} for $R_\mathrm{th} =$ \SI{0.271}{\nano\meter} and \SI{0.3}{\nano\meter}. At higher SoC, particularly SoC \SI{70}{\%}--\SI{80}{\%}, the static and dynamic results become similar again because the time-dependent SoC decrease remains within a relatively flat region of the Birkl SoC--OCV curve. Although the transition times remain similar, in the dynamic $\mathrm{SoC}_0=\SI{80}{\%}$ case, the decrease in SoC reduces the supply of $\ce{Li^0}$ radicals and consequently modifies surface relaxation relative to the corresponding static-SoC case. The regime changes induced by dynamic SoC reflect the SoC--OCV relation across the investigated SoC range. In regions where the OCV changes substantially with SoC, the reduced $\ce{Li^0}$ radicals supply limits the contribution of surface relaxation and increases the influence of noise-induced roughening. This implies the regime shifts observed in Figure~\ref{fig:transition_time}(e) and (f), indicating that the regime boundaries are not determined by the initial SoC alone.

From an engineering perspective, the earlier transition observed under dynamic SoC conditions indicates that the static SoC assumption can overestimate the stability of dense SEI in the SoC \SI{60}{\%}--\SI{65}{\%} range. This occurs because the time-dependent SoC reduction moves the system into the transitional region of the SoC--OCV curve near SoC \SI{60}{\%}. In this region, even a small reduction in SoC produces a pronounced increase in OCV, so the SEI can reach the porous state earlier than predicted under static conditions. Therefore, long-term SEI stability under open-circuit storage should be evaluated by considering not only the initial SoC, but also its time-dependent decrease caused by irreversible capacity loss. By coupling the SoC--OCV relation with a capacity-loss model, the present framework enables both static and dynamic dense-to-porous transition patterns to be evaluated consistently. This framework can be extended to other battery systems when the corresponding SoC--OCV relation and capacity-loss model are available, together with appropriate recalibration of the relevant transport and reaction parameters.

    \section{Conclusion}
This study investigated the dense-to-porous transition of the SEI under open-circuit conditions using a phase-field model that couples transport-limited SEI growth with random noise flux. The model reproduced the characteristic $\sqrt{t}$ dependence of SEI growth over the investigated SoC range, providing a baseline for analyzing the evolution of interfacial roughness. The analysis of different noise fields showed that the correlation length affects both the magnitude and the evolution pattern of the interface roughness. A correlation length of $L_\mathrm{c}=\SI{2}{nm}$ was selected because it resembles the typical size of grains inside the SEI layer.

Under static SoC conditions, the interface exhibited distinct roughness patterns over different SoC ranges, which were described by three regimes. These regimes reflect changes between noise-induced roughening and surface relaxation supported by the supply of $\ce{Li^0}$ radicals. Since the interfacial supply of $\ce{Li^0}$ radicals depends on the OCV, this balance is inherently linked to the SoC--OCV relation. The variation in the graphite SoC--OCV relation therefore provides an electrochemical basis for the SoC-dependent dense-to-porous transition. This relationship is not specific to the graphite anode, as the SoC--OCV profile is expected to play an important role in determining the dense-to-porous transition behavior in other battery systems as well. To quantify this SoC-dependent transition behavior, physically motivated interface roughness thresholds of \SI{0.271}{nm} and \SI{0.3}{nm} were introduced, revealing that the transition develops gradually rather than as an instantaneous morphological change. The transition time generally increased with SoC, with a pronounced change near SoC \SI{55}{\%}--\SI{60}{\%}, where the transition between the identified roughness regimes was observed.

The impact of irreversible capacity loss on SoC further showed that the dense-to-porous transition cannot be determined from the initial static SoC. Dynamic SoC conditions modified the roughness evolution and shifted the transitional behavior between regimes to higher initial SoC values. The effect of dynamic SoC on the roughness regime depends strongly on the SoC--OCV profile: the regime is maintained within relatively flat regions, whereas a regime change can occur when the SoC evolves through a region where the OCV changes rapidly. These results demonstrate that the long-term morphological evolution of the SEI depends on both the initial electrochemical state and its time-dependent capacity during storage. The proposed framework provides a basis for evaluating SoC-dependent SEI stability under both static and dynamic open-circuit conditions. The dynamic SoC analysis further confirms the importance of the SoC--OCV relation identified under static conditions, showing that the OCV profile plays a central role in determining how the dense-to-porous transition time evolves with SoC.
    \section{Acknowledgement}
J.J. and I.E.C. acknowledge support from the Independent Research Fund Denmark (Green Transition Project 1, project “Reconfigurable Metamaterials for Next Generation High-capacity Batteries” under grant number 0217-00111B). The authors acknowledges support from the Novo Nordisk Foundation Data Science Research Infrastructure 2022 Grant:  A high-performance computing infrastructure for data-driven research on sustainable energy materials, Grant no. NNF22OC0078009. Furthermore, the authors acknowledge support from the European Union’s Horizon Europe research and innovation programme under grant agreements No. 101137725 and 101103997 (BatCAT and DigiBatt). This work contributes to the research performed at CELEST (Center for Electrochemical Energy Storage Ulm-Karlsruhe).

    % \input{5discussion}
    % \clearpage % can remove if you don't want a new page
    % \input{5discussion}
    % \clearpage % can remove if you don't want a new page
    % \input{6methods}

    % \begin{singlespace}
    %     \printbibliography % for biblatex
    %     %\bibliography{library} % for natbib
    % \end{singlespace}
    % \clearpage

    % \section*{Supplementary Information}
    % \setcounter{page}{1} % Set page at 1
    % \input{supplement/0suppl_text}
    % \input{supplement/1suppl_figs}
    % \input{supplement/2suppl_tabs}
    
    % \begin{figure}[!htb]
    %     \centering
    %     \includegraphics[width=3.33in]{"figs/tortuosity_figure.pdf"}
    %     \caption{Schematics of surfaces with different dendrite height and tortuosity. (a) height = \SI{20}{\micro\meter}, tortuosity = 1.08, (b) height = \SI{5}{\micro\meter}, tortuosity = 1.94, (c) height = \SI{20}{\micro\meter}, tortuosity = 1.16, (d) height = \SI{20}{\micro\meter}, tortuosity = 4.26.}
    %     \label{fig:height_tortuosity}
    % \end{figure}
\clearpage
\printbibliography % for biblatex
\clearpage
\renewcommand{\thesubsection}{S\arabic{section}.\arabic{subsection}}
\renewcommand{\thefigure}{S\arabic{figure}}
\renewcommand{\thetable}{S\arabic{table}}
\renewcommand{\theequation}{S\arabic{equation}}

\setcounter{section}{0}
\setcounter{figure}{0}
\setcounter{table}{0}
\setcounter{equation}{0}

\section{Supplementary Information}
\begin{figure}[!ht]
    \centering
    \includegraphics[width=150 mm]{"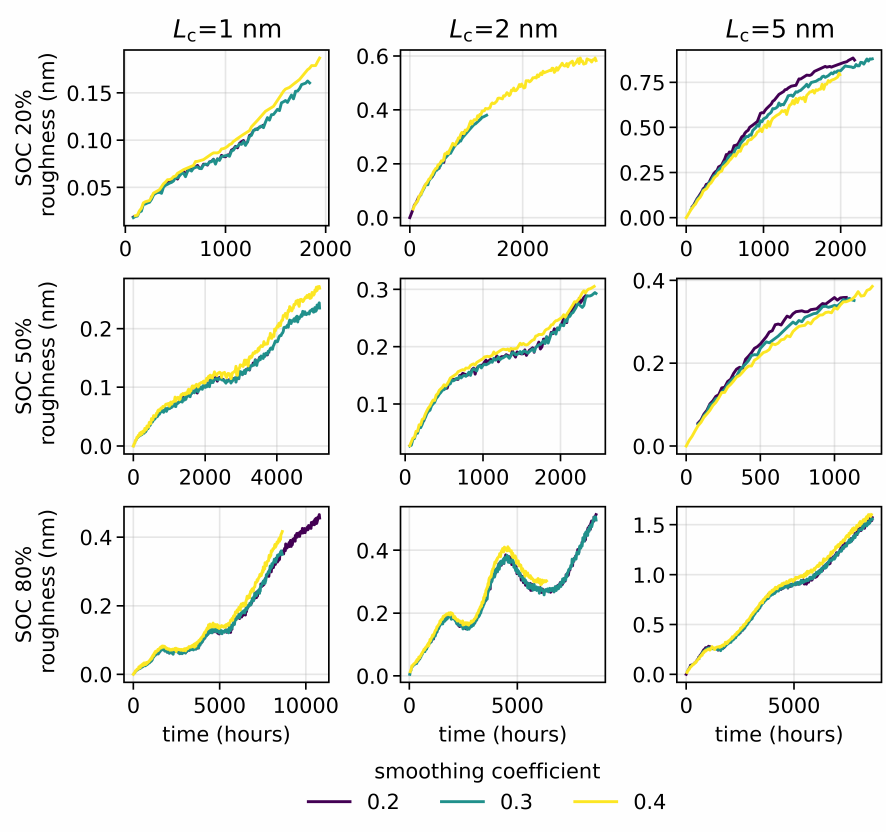"}
    \caption{Effect of the smoothing coefficient on the interface roughness evolution. Each panel shows the roughness as a function of time for smoothing coefficients of $s = 0.2$, $0.3$, and $0.4$. Columns correspond to correlation lengths of $L_\mathrm{c} = 1$, $2$, and \SI{5}{nm}, and rows to SoC \SI{20}{\%}, \SI{50}{\%}, and \SI{80}{\%}. The three curves nearly overlap in every panel, indicating that the smoothing coefficient has little effect on the roughness evolution over the range.}
    \label{fig:noise_effect_smoothing_factor}
\end{figure}
\end{document}